\documentclass[reprint,aps,prc,twocolumn,superscriptaddress,floatfix,amsmath,amssymb,10pt]{revtex4-2}

 \usepackage{graphicx}
\usepackage{bm}
\usepackage{amsmath} \usepackage{braket} \usepackage{epsfig}
\usepackage{tensor}
\usepackage{multirow}
\usepackage[T1]{fontenc}
\usepackage[version=4]{mhchem}
\usepackage{titlesec}
\usepackage{ifthen}
\usepackage{sidecap}
\usepackage{listings}
\usepackage[para,online,flushleft]{threeparttablex}
\usepackage{algorithm}
\usepackage{algpseudocode}
\usepackage{comment}
\usepackage{MnSymbol}
\usepackage{xr-hyper}
\usepackage{hyperref}
\hypersetup{breaklinks=true,colorlinks=true,linkcolor=blue,citecolor=blue,filecolor=magenta,urlcolor=cyan}

\usepackage{orcidlink}

\renewcommand{\vec}[1]{\mbox{\boldmath $#1$}}

\usepackage[all]{hypcap}

\renewcommand{\vec}[1]{\mbox{\boldmath $#1$}}
\newcommand{\PRLsep}{\noindent\makebox[\linewidth]{\resizebox{0.3333\linewidth}{1pt}{$\bullet$}}\bigskip}
\newcommand{\subref}[2]{\hyperref[#1]{\ref{#1}(#2)}}

\newcommand{\rpa}{$r$-process}
\newcommand{\rpn}{$r$ process}

\usepackage[normalem]{ulem}

\usepackage{xcolor}
\definecolor{pastelgray}{rgb}{0.81, 0.81, 0.77}
\definecolor{beaublue}{rgb}{0.9, 0.9, 0.93}

\begin{document}
\title{Multimodal fission from self-consistent calculations}

\author{Daniel Lay\orcidlink{0000-0001-8947-391X}}
\affiliation{Department of Physics and Astronomy and FRIB Laboratory, Michigan State University, East Lansing, Michigan 48824, USA}

\author{Eric Flynn\orcidlink{0000-0002-6641-3620}}
\affiliation{Department of Physics and Astronomy and FRIB Laboratory, Michigan State University, East Lansing, Michigan 48824, USA}

\author{Sylvester Agbemava\orcidlink{0000-0003-2432-2402}}
\affiliation{FRIB Laboratory, Michigan State University, East Lansing, Michigan 48824, USA}

\author{Kyle Godbey\orcidlink{0000-0003-0622-3646}}
\affiliation{FRIB Laboratory, Michigan State University, East Lansing, Michigan 48824, USA}

\author{Witold Nazarewicz\orcidlink{0000-0002-8084-7425}}
\affiliation{Department of Physics and Astronomy and FRIB Laboratory, Michigan State University, East Lansing, Michigan 48824, USA}

\author{Samuel A. Giuliani\orcidlink{0000-0002-9814-0719}}
\affiliation{Departamento de F{\'i}sica Te{\'o}rica and CIAFF, Universidad Aut{\'o}noma de Madrid, Madrid 28049, Spain}
\affiliation{Department of Physics, Faculty of Engineering and Physical Sciences, University of Surrey, Guildford, Surrey GU2 7XH, United Kingdom.}

\author{Jhilam Sadhukhan\orcidlink{0000-0003-1963-1390}}
\affiliation{Physics Group, Variable Energy Cyclotron Centre, Kolkata 700064, India }

\begin{abstract}
\edef\oldrightskip{\the\rightskip}
\begin{description}
\rightskip\oldrightskip\relax
\setlength{\parskip}{0pt}

\item[Background]
When multiple fission modes coexist in a given nucleus, 
distinct fragment yield distributions appear. Multimodal fission has been observed in a number of fissioning nuclei spanning the nuclear chart, and this phenomenon is expected to affect the nuclear abundances synthesized during the rapid neutron-capture process ({\rpa}).
\item[Purpose]
In this study, we generalize the previously proposed hybrid model for fission-fragment yield distributions to predict competing fission modes and estimate the resulting yield distributions. Our framework allows for a comprehensive large-scale calculation of fission fragment yields suited for {\rpa} nuclear network studies. 
\item[Methods]
Nuclear density functional theory is employed to obtain the potential energy and collective inertia tensor on a multidimensional collective space defined by mass multipole moments. Fission pathways and their relative probabilities are determined using the nudged elastic band method. Based on this information, mass and charge fission yields are predicted using the recently developed hybrid model. 
\item[Results]
Fission properties of fermium isotopes are calculated in the axial quadrupole-octupole collective space for three energy density functionals (EDFs). Disagreement between the EDFs appears when multiple fission modes are present. Within our framework, the UNEDF1$_{\textrm{HFB}}$ EDF agrees best with experimental data. Calculations in the axial quadrupole-octupole-hexadecapole collective space improve the agreement with the experiment for  SkM$^*$. We also discuss the sensitivity of fission predictions on the choice of EDF  for several superheavy nuclei.
\item[Conclusions] 
Fission fragment yield predictions for nuclei with multiple fission modes are sensitive to the underlying EDF. For large-scale calculations in which a minimal number of collective coordinates is considered,  UNEDF1$_\textrm{HFB}$ provides the best description of experimental data, though the sensitivity motivates robust quantification of the uncertainties of the theoretical model.

\end{description}
\end{abstract}

\date{\today}
\maketitle

\section{Introduction}
Nuclear fission is expected to play a major role during the rapid neutron-capture nucleosynthesis process (\rpn)~\cite{Burbidge1957,Cameron1957} that occurs in astrophysical scenarios with sufficiently high neutron densities~\cite{Horowitz2018,Kajino2019,Cowan2021}. Due to the occurrence of fission cycling, fragment yields of neutron-rich fissioning nuclei are predicted to shape the \rpa\ abundances in the $110 \lesssim A \lesssim 170$ region~\cite{Steinberg1978a,Rauscher1994,Panov2001,Panov2008,Eichler2015,Goriely2015,Vassh2018,Vassh2019,Lemaitre2021,Panov2021,Panov2023,Kullmann2023}. Additionally, multimodal fission has been argued to impact the shape of the rare-earth peak of the \rpa\ abundance distributions~\cite{Goriely2013}. Therefore, the ability to efficiently identify possible fission pathways is important for global calculations of {\rpa} nuclei.

\begin{figure}[tbh]
    \includegraphics[width=\linewidth]{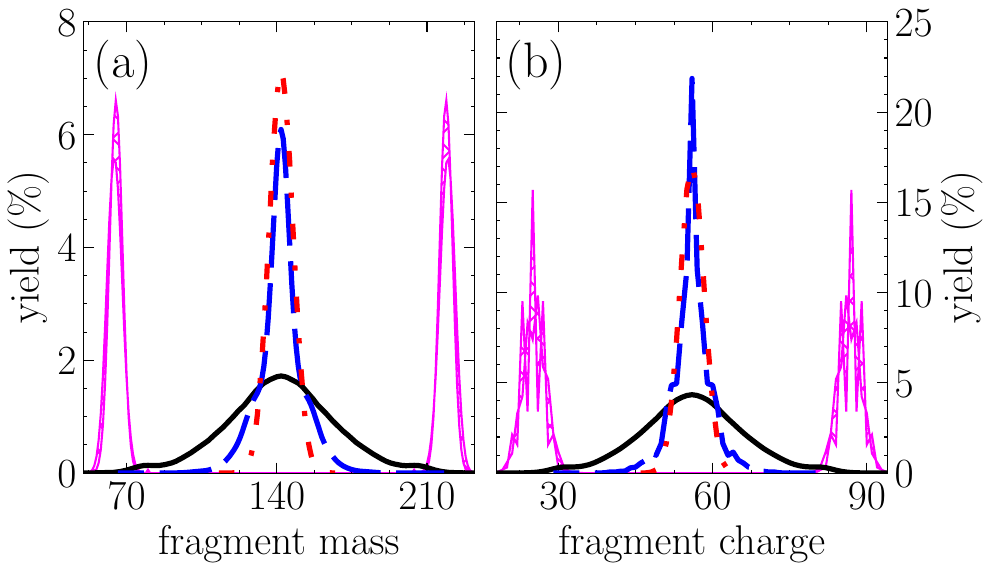}    
    \caption{(a) Mass and (b) charge yield distributions of $^{284}$Cn. The magenta patterned regions are calculated with the UNEDF1$_\textrm{HFB}$ functional (our model). Also shown are the results from the BSM~\cite{Mumpower2020} (black solid lines), SPM~\cite{Lemaitre2021} (blue dashed lines), and GEF~\cite{SCHMIDT2016} (red dash-dotted lines) models.}
    \label{fig:284Cn-yields}
\end{figure}

Since most of the nuclei synthesized during the \rpa\ neutron irradiation phase cannot be measured, missing data for nuclear reaction network simulations must be obtained theoretically. However, predictions of current models for fragment yields of neutron-rich fissioning nuclei are not consistent. As an example, Fig.~\ref{fig:284Cn-yields} shows the fragment yields for $^{284}$Cn computed using four models. The framework discussed in this paper predicts strongly asymmetric fission (cluster decay), while the other models predict symmetric distributions of varying widths.

Multimodal fission is found in regions of the nuclear chart between nuclei with symmetric and asymmetric fission modes. Examples of nuclei that show competing fission modes are:   $^{257,258}$Fm, $^{260}$Md, and $^{258}$No~\cite{Balagna1971, John1971, Hulet1986, Hulet1989}. The coexistence of fission pathways is caused by the delicate balance between competing shell effects and nuclear dynamics, providing a stringent test of fission models. Much theoretical progress has been made in understanding multimodal fission. A number of studies have used macroscopic-microscopic approaches, either with~\cite{Usang2017,Miyamoto2019,Albertsson2021} or without~\cite{Moller1987,Ichikawa2009} inclusion of collective dynamics. Other studies have used self-consistent density functional theory (DFT)~\cite{Warda2002,Dubray2008,Staszczak2009}, although fragment yields were not reported. Investigations using the time-dependent generator coordinate method have also been carried out~\cite{Regnier2019}.

In this work, we use constrained nuclear DFT and the nudged elastic band method~\cite{OurNEBPaper} to efficiently identify competing fission pathways for fermium isotopes and selected superheavy nuclei. We then use the hybrid approach proposed in~Refs.~\cite{Sadhukhan2020,Sadhukhan2022} to compute spontaneous fission (SF) fragment yields. We also study the sensitivity of predictions on the size of the collective space and the choice of EDF. 

The paper is organized as follows: Section~\ref{sec:model} reviews the methodology used to compute the fragment yields. Sections~\ref{sec:results2D} and \ref{sec:results3D} discuss the results obtained for quadrupole-octupole and quadrupole-octupole-hexadecapole collective spaces, respectively. Conclusions and prospects are summarized in Sec.~\ref{sec:conclusions}.

\section{Model}
\label{sec:model}
Potential energy surfaces (PESs) and the collective inertia tensor are calculated within the nuclear DFT framework~\cite{BenderRMP}. We consider two Skyrme EDFs: SkM$^*$~\cite{BARTEL198279} and UNEDF1$_\textrm{HFB}$~\cite{Schunck_2015}, in the particle-hole channel. The SkM$^*$ EDF is a standard for fission calculations, while UNEDF1$_\textrm{HFB}$ extends the optimization protocol of UNEDF0~\cite{Kortelainen2010} to include the additional data on fission isomers; it describes pairing correlations at the Hartree-Fock-Boguliubov (HFB) level. The pairing EDF is approximated using the mixed-type density-dependent delta interaction~\cite{DobaczewskiPairing}. We also consider the D1S parametrization~\cite{berger1984microscopic} of the Gogny interaction, which has been used for studying fission properties of heavy and superheavy nuclei~\cite{Robledo2018}.  We restrict our calculations to axially symmetric shapes, as these are sufficient to describe the outer turning surface defining fission fragment yields. Specifically, we constrain the mass quadrupole ($Q_{20}$), octupole ($Q_{30}$), and hexadecapole ($Q_{40}$) moments, defined as
\begin{align}
    {Q}_{\lambda 0}  =  \int d^{3}r\rho{(\vec{r})}r^\lambda Y_{\lambda 0},
\end{align}
where $\rho{(\vec{r})}$ is the total nucleonic density at $\vec{r}$ and $Y_{\lambda 0}$ are spherical harmonics. To evaluate action integrals, we introduce dimensionless collective coordinates~\cite{Sadhukhan2014}.

To solve the HFB equations, we employ the axial DFT solver HFBTHO (v3.00)~\cite{PEREZ2017363} for the UNEDF1$_\textrm{HFB}$ and SkM$^{*}$ EDFs, and HFBaxial~\cite{Robledo2011d} for the D1S EDF. These solvers use the harmonic oscillator basis expansion. To ensure convergence, we set the maximum oscillator number of the harmonic oscillator basis to 25.

The collective inertia tensor is calculated in the adiabatic time-dependent HFB framework, employing the perturbative cranking approximation~\cite{Yuldashbaeva1999, Baran2011, Giuliani2018b}. Given the number of configurations explored, this approach to the inertia tensor enhances computational efficiency while maintaining reasonable numerical accuracy. For simplicity, we neglect the dynamical pairing degree of freedom~\cite{Moretto1974,Sadhukhan2014,Giuliani2014,Zhao2016} as we do not carry out a detailed SF lifetime analysis. 

Fission pathways are equivalent to least action paths (LAPs) obtained by minimizing the collective action integral~\cite{Brack1972}:
\begin{equation}
\label{action-integral}
S(L)=\frac{1}{\hbar} \int_{s_{\rm in}}^{s_{\rm out}} \sqrt{2\mathcal{M}_{\text{eff}}(s)\left(V_{\text{eff}}(s)- \Delta E\right)}\,ds \,,
\end{equation}
where $V_{\text{eff}}(s)$ and
$\mathcal{M}_{\text{eff}}(s)$ are the effective potential energy and collective inertia, respectively, along the path $L(s)$.  The integration limits $s_{\rm in}$ and $s_{\rm out}$ correspond to the classical inner and outer turning points, respectively, determined by the condition $V_{\text{eff}}(s) = \Delta E$. The quantity $\Delta E$ represents the zero-point correction to the ground-state energy. Throughout this work, following discussion in Ref.~\cite{OurNEBPaper}, we refer to the contour of energy $V_\textrm{eff}=\Delta E$ as the outer turning line, and the collective coordinates at $s_\textrm{out}$ as the exit point. We use the nudged elastic band method (NEB)~\cite{OurNEBPaper} to compute LAPs, as this method is capable of identifying LAPs corresponding to different fission modes. The supplemental material (SM)~\cite{SM} provides further details on the PES and LAP calculations.

If $S_i$ is the action along the $i$'th competing fission path, the relative normalized probability of this mode is
\begin{equation}
P_i\approx\frac{\exp(-2S_i)}{\sum_{j=1,n}\exp(-2S_j)}.
\label{relative_P}
\end{equation}

The parameter $\Delta E$ can be estimated by means of the generator coordinate method~\cite{Staszczak1989,BARAN2015,schunck2016}. However, $\Delta E$ is often treated as a free parameter that can be adjusted to lifetime measurements~\cite{Staszczak2011, Giuliani2013,Sadhukhan2016}. Here, $\Delta E$ is only used to determine the prescission configuration $\mathcal{C}$ with the expectation that the fission fragment yields estimated at $\mathcal{C}$ are independent of $\Delta E$. We demonstrate $\Delta E$ independence in SM~\cite{SM}, and we take $\mathcal{C}$ to be the exit point configuration computed with $\Delta E=0$.

Fission isomers (FIs) are local minima in the PES at deformations between the g.s. and the outer turning line. In some transactinide nuclei, FIs can have lower potential energy than the g.s. configuration. If $\Delta E$ is chosen relative to the g.s., the FI may be classically accessible. However, Eq.~(\ref{action-integral}) assumes a tunneling configuration. To assess the influence of the FI on the fragment yields, we set $\Delta E=0$ relative to the configuration with lower energy. While in some cases, LAPs starting from the g.s. may bypass the FI~\cite{Sadhukhan2014}, this does not occur for the cases considered in this work. So, via Eq.~(\ref{relative_P}), $P_i$ is independent of the action calculated in between g.s. and FI.
\begin{figure*}[htb]
    \includegraphics[width=0.8\linewidth]{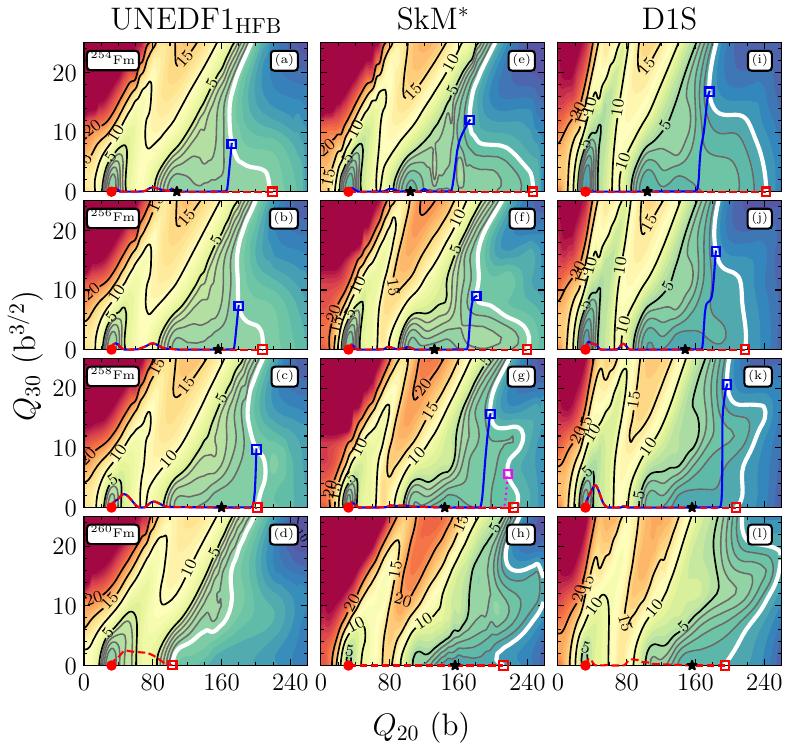}
    \caption{Potential energy surfaces of $^{254,256,258,260}$Fm (in MeV) calculated using UNEDF1$_\textrm{HFB}$ (left), SkM$^*$ (center), and D1S  (right) EDFs. The symmetric (dashed lines) and asymmetric (solid and dotted lines) least action paths are drawn from the ground state (filled circle) to the fission isomer (asterisk), to the exit point (open square). The white contour denotes the outer turning line $V_\textrm{eff}=\Delta E =0$. Gray contours are marked at 1 MeV intervals for $0<V_\textrm{eff}<5$ MeV.}
    \label{fig:2d-Fm-surfaces}
\end{figure*}

Once the prescission configuration $\mathcal{C}$ has been determined, the fission fragment yield is computed using the hybrid method developed in~\cite{Sadhukhan2020, Sadhukhan2022}. This approach identifies prefragments using the nucleonic localization functions (NLFs) at $\mathcal{C}$. The remaining nucleons, the so-called neck nucleons, are distributed between the two prefragments according to a microcanonical probability distribution. For details on the prefragment selection, see SM~\cite{SM} and ~\cite{Sadhukhan2017}. The combined fragment yields are computed by taking the weighted sum of the yields and their corresponding $P_i$.

\section{Multimodal fission in 2D}\label{sec:results2D}
In this section, we investigate fission in the two-dimensional collective space $(Q_{20},Q_{30})$. We first study the benchmark chain of Fm isotopes, which is known to transition from asymmetric to symmetric fission with increasing neutron number~\cite{Hulet1989,ITKIS2015}. Next, we investigate selected superheavy nuclei which are expected to show stronger model dependence because of their reduced fission barriers.

\subsection{Fermium isotopes}
\label{subsec:2d-Fm}
The even-even Fm isotopes with $154\le N\le160$ are known to undergo a transition from asymmetric fission characteristic of lighter actinides to symmetric fission as $N$ increases towards $N=164$. This transition is related to strong shell effects in prefragments as they approach the doubly magic nucleus $^{132}$Sn. The resulting bimodal fission in intermediate isotopes was investigated in numerous papers~\cite{Depta1986,Moller1987,Moller1989,Moller2001,Warda2002,warda2003,Warda2004,asano2004dynamical,Bonneau2006,Staszczak2009,Simenel2014,Scamps2015,Miyamoto2019,Regnier2019,Albertsson2021,Sadhukhan2020,Sadhukhan2022,pomorski2023fission,bernard2023}. We note that the PESs computed in Ref.~\cite{Sadhukhan2020} used the triaxial solver HFODD~\cite{SCHUNCK2017145} with the constraint $Q_{22}=0$, on a smaller $(Q_{20},Q_{30})$ grid. 

Figure~\ref{fig:2d-Fm-surfaces} shows the PESs and LAPs calculated for $^{254,256,258,260}$Fm with different EDFs. In general, the topologies of PESs agree well between various EDFs. The fission barriers obtained with SkM$^*$ and D1S are higher than those of UNEDF1$_\textrm{HFB}$. Except for $^{260}$Fm with UNEDF1$_\textrm{HFB}$ (Fig.~\ref{fig:2d-Fm-surfaces}(d)), all the PESs predict a FI at $100<Q_{20}<160$\,b and $Q_{30}=0$. 

\begin{figure}[htb]
\includegraphics[width=0.9\linewidth]{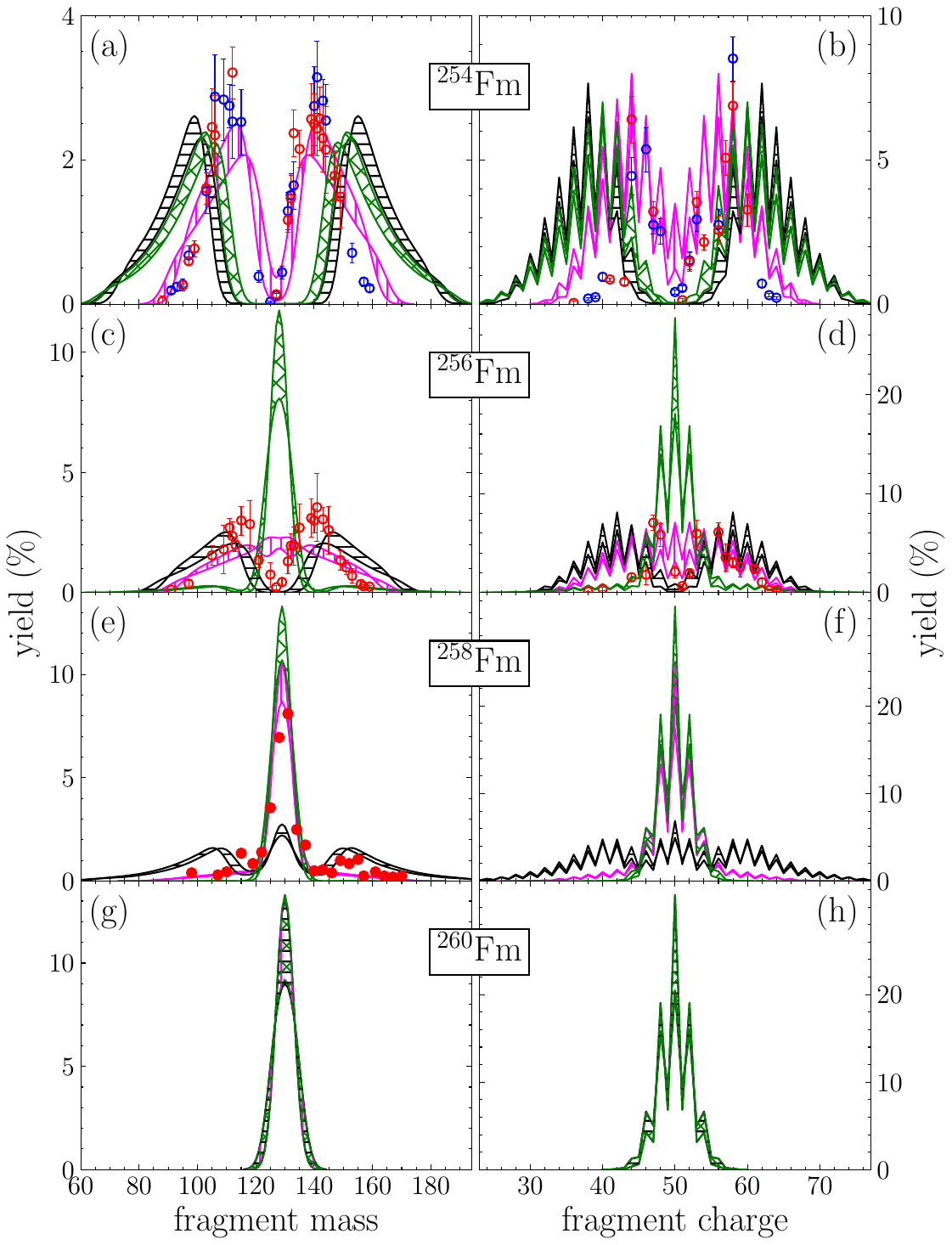} 
\caption{Fission fragment mass (left) and charge (right) yields for $^{254,256,258,260}$Fm calculated with UNEDF1$_\textrm{HFB}$ (magenta vertical patterns), SkM$^*$ (black horizontal patterns), and D1S (green $\times$ patterns). Experimental yields (circles)~\cite{PhysRevC.8.1488,PhysRevC.16.1483,PhysRevC.5.1725,PhysRevC.21.972} are shown where available.}
    \label{fig:2d-fm-yields}
\end{figure}

Since triaxiality is ignored in this study, in most cases, one path along $Q_{30}\approx0$ connects the g.s. and the FI. The reflection-symmetric path continues past the FI, and then a bifurcation resulting in a coexistence of symmetric and asymmetric pathways takes place. The asymmetric paths are smooth at the bifurcation point. Some path segments are straight due to the flatness of the PES. For $^{258}$Fm calculated with SkM$^*$, Fig.~\ref{fig:2d-Fm-surfaces}(g), we observe an additional asymmetric LAP that ends close to the symmetry axis, due to the geometry of the outer turning surface. However, within our hybrid approach, this exit point corresponds to a wide symmetric yield distribution.

The relative probability $P_s$ of the symmetric mode is given in Table~\ref{table:2d-rel_action}. All EDFs transition from asymmetric-dominant to symmetric-dominant fission path with increasing $N$.  Competition between the modes is expected in $^{256,258}$Fm for UNEDF1$_\textrm{HFB}$,
$^{258}$Fm for SkM$^*$, and $^{256}$Fm for D1S.
The competition between symmetric and asymmetric fission modes seen in Table~\ref{table:2d-rel_action} is generally consistent
with SkM$^*$ results of Refs.~\cite{Staszczak2009,Staszczak2011}.
The small differences are most likely due to (i) our inclusion of the multidimensional collective inertia tensor in computing the collective action and (ii) our precise minimization of $S(L)$.

\begin{table}[tb]
    \centering
    \caption{The relative probability $P_s$ of the symmetric mode for the 2D calculation of Fm isotopes. }
    \begin{ruledtabular}
    \begin{tabular}{lcccc}
                   & $^{254}$Fm & $^{256}$Fm & $^{258}$Fm & $^{260}$Fm  \\
         \hline
         UNEDF1$_\textrm{HFB}$ & $\sim$0 & $0.12$ & $0.79$ & $1$ \\
         SkM$^{*}$ & $\sim$0 & $\sim$0 & $0.17$ & $ 1$ \\
         D1S  & $\sim$0 & $0.88$ & $1$ & $1$  \\
    \end{tabular}
    \end{ruledtabular}
    \label{table:2d-rel_action}
\end{table}
Figure~\ref{fig:2d-fm-yields} shows the calculated yields, together with experimental data~\cite{PhysRevC.8.1488,PhysRevC.16.1483,PhysRevC.5.1725,PhysRevC.21.972}. The experimental data for $^{254,256}$Fm show an asymmetric distribution, while for $^{258}$Fm, the yields are primarily symmetric, with a small asymmetric shoulder. The error bands in the calculated distributions originate from the two-particle uncertainty~\cite{Sadhukhan2020}. Our UNEDF1$_\textrm{HFB}$ results are in close agreement with the data for $^{254}$Fm. Competition between modes for $^{256,258}$Fm is present for each EDF, although the D1S model overestimates the symmetric contribution. All EDFs predict overlapping symmetric yields for $^{260}$Fm. The transition from asymmetric to symmetric fission is clearly present, albeit at different neutron numbers for the EDFs. Despite the overestimation of the symmetric mode for $^{256}$Fm, UNEDF1$_\textrm{HFB}$ provides the best description overall.

\begin{figure}[htb]
    \includegraphics[width=\linewidth]{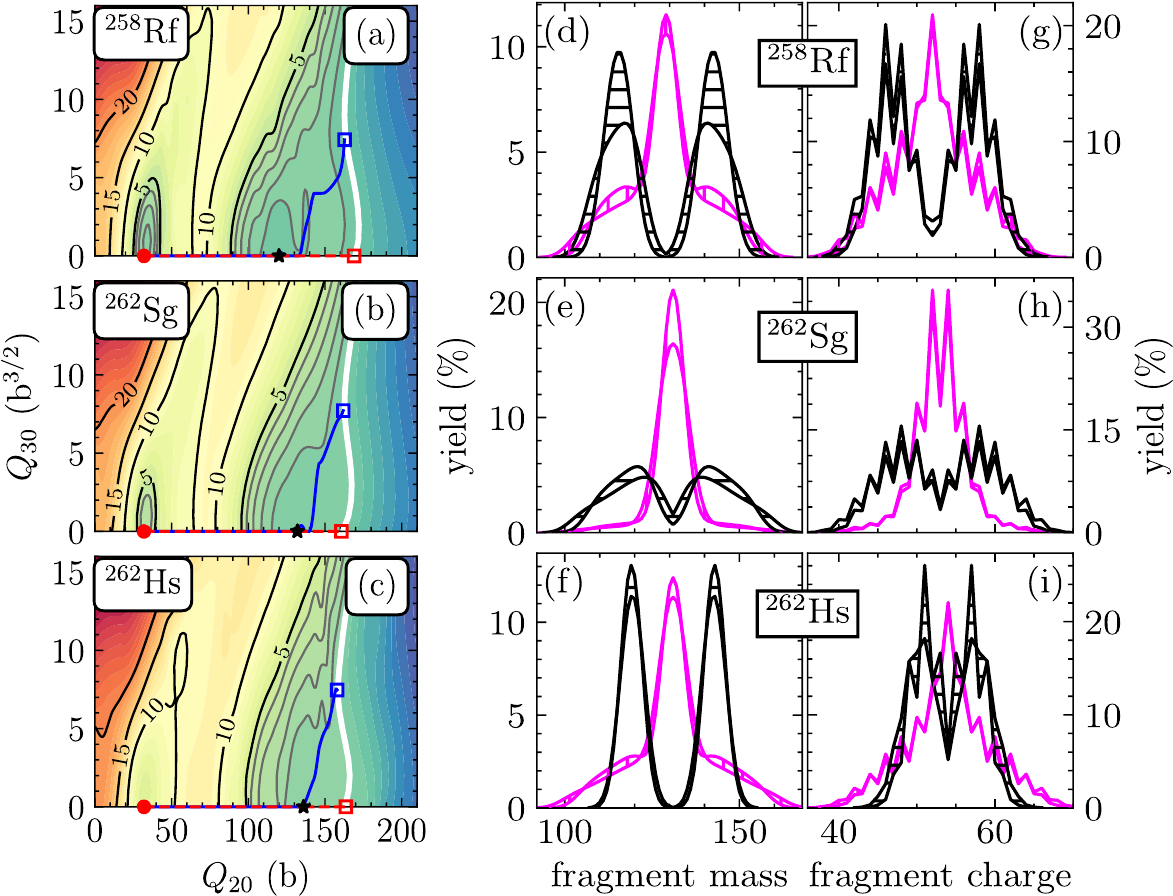} 
    \caption{Panels (a)-(c): similar as in Fig.~\ref{fig:2d-Fm-surfaces} but for $^{258}$Rf, $^{262}$Sg, and $^{262}$Hs using UNEDF1$_\textrm{HFB}$. Panels (d)-(f) and (g)-(i):  mass and charge yields, respectively, for UNEDF1$_\textrm{HFB}$ (magenta vertical patterns) and SkM$^*$ (black horizontal patterns).} 
    \label{fig:2d-SH-yields}
\end{figure}

\begin{figure*}[htb]
    \includegraphics[width=1.0\linewidth]{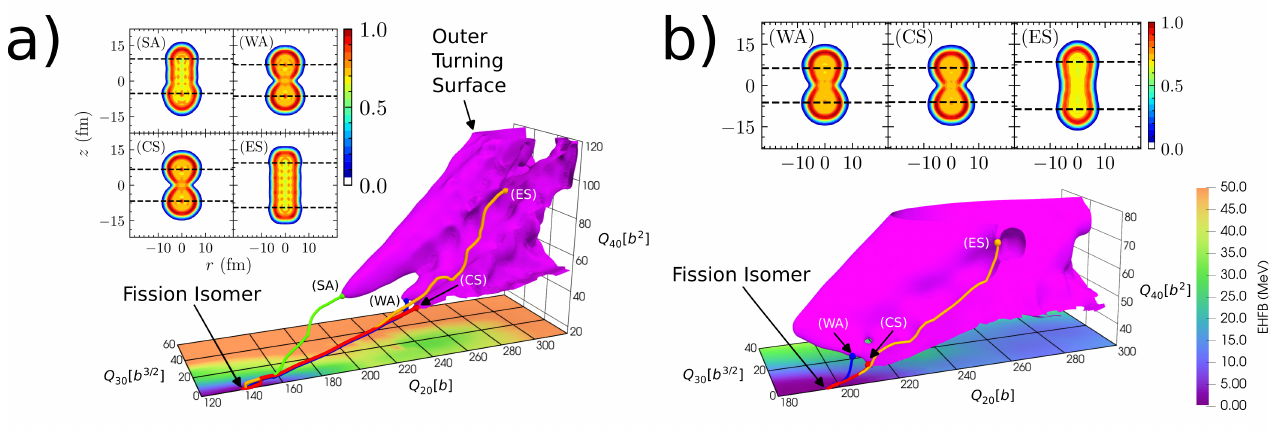}    
    \caption{The least action paths in 3D for $^{258}$Fm using SkM$^*$ (a) and UNEDF1$_\textrm{HFB}$ (b). The outer turning surface is shown. A 2D PES is shown for constant $Q_{40} = 16$~b$^{2}$ ($Q_{40}=32$~b$^2$) for SkM$^{*}$ (UNEDF1$_\textrm{HFB}$). Neutron localizations for the identified precission configurations are shown in the insets.}
    \label{fig:3d-Fm-surface}
\end{figure*}
\subsection{Superheavy nuclei}
The superheavy nuclei $(Z>104)$ are stabilized only by quantum shell effects~\cite{Smolanczuk1995} and hence they  are short-lived. The dominant decay modes observed in this region are $\alpha$ decay and SF~\cite{somerville1985,SOBICZEWSKI2006,gregorich2006,gates2008,ITKIS2015,GiulianiRMP}. Some isotopes of Fl and Ts are observed to undergo SF after several $\alpha$ emissions~\cite{Khuyagbaatar2014,Utyonkov2018}. Neutron-rich superheavy nuclei may decay directly via SF~\cite{Staszczak2013,Goriely2015,SCHMIDT2016,AFANASJEV2018,Giuliani2018,Mumpower2020,Lemaitre2021} marking the end of the \rpa~path. Multimodal fission has been studied in the superheavy region in~Refs.\,\cite{Staszczak2009,BARAN2015}.

The PESs and the corresponding symmetric and asymmetric LAPs computed with UNEDF1$_\textrm{HFB}$ for $^{258}$Rf, $^{262}$Sg, and $^{262}$Hs  are shown in Fig.~\ref{fig:2d-SH-yields}(a-c).  For D1S, the asymmetric mode is not relevant for $^{262}$Sg. Consequently, the remaining discussion pertains to  SkM$^*$ and UNEDF1$_\textrm{HFB}$ only.

\begin{table}[htb]
    \caption{Similar as Table~\ref{table:2d-rel_action} except for $^{258}$Rf, $^{262}$Sg and $^{262}$Hs.}
    \begin{ruledtabular}
    \begin{tabular}{lccc}
                   & $^{258}$Rf & $^{262}$Sg & $^{262}$Hs\\
         \hline
         UNEDF1$_\textrm{HFB}$ & $0.48$ & $0.85$ & $0.52$\\
         SkM$^{*}$ & $\sim$0 & $0.04$ & $\sim$0\\
    \end{tabular}
    \end{ruledtabular}
    \label{table:SH-2D-rel_action}
\end{table}
Table~\ref{table:SH-2D-rel_action} shows $P_s$ for the superheavy nuclei shown in Fig.~\ref{fig:2d-SH-yields}. SkM$^*$ favors the asymmetric mode, while UNEDF1$_\textrm{HFB}$ predicts competition between the modes. The modes predicted by SkM$^*$ are consistent with those identified in~\cite{Staszczak2009}, and the reasons for small differences in $P_s$ were discussed in Sec.~\ref{subsec:2d-Fm}.

The mass and charge yields are shown in Fig.~\ref{fig:2d-SH-yields}(d-i). Similar to $^{258}$Fm, the yields computed using UNEDF1$_\textrm{HFB}$ have a large symmetric peak with asymmetric shoulders. Note that the fragment yield peak for each mode agrees between EDFs, but contributes differently, according to its relative probability. SPM~\cite{Lemaitre2021} and GEF~\cite{SCHMIDT2016} predict asymmetric yields for $^{258}$Rf. BSM~\cite{Mumpower2020} predicts asymmetric yields for $^{258}$Rf and $^{262}$Sg, and symmetric yields for $^{262}$Hs. Symmetric fission is predicted in~\cite{BARAN2015} for the three nuclei, although yields are not computed. The large spread between theoretical predictions highlights the usefulness of multimodal fission for differentiating between models.

\section{Multimodal fission in 3D}\label{sec:results3D}
In order to separate fission pathways that are unresolved in a lower-dimensional collective space, it is advisable to consider additional collective degrees of freedom such as the hexadecapole moment $Q_{40}$ that  controls the number of nucleons in the neck region. Indeed, studies constraining $Q_{40}$ have identified fission modes not found in the reduced $(Q_{20},Q_{30})$ collective space~\cite{berger1984microscopic,berger1989,Warda2002,Staszczak2009,TSEKHANOVICH2019,schunck2016,Zdeb2021,Chi2023}. In this section, we study fission in the three-dimensional collective space $(Q_{20},Q_{30},Q_{40})$. 

\begin{figure}[htb]
    \includegraphics[width=\linewidth]{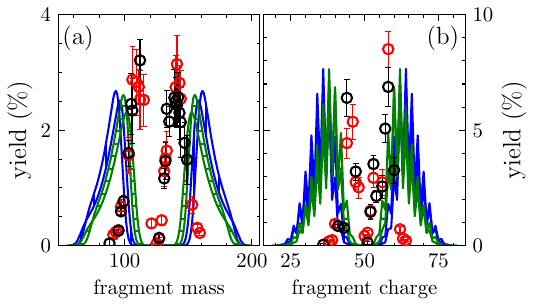}
    \caption{The fragment mass (a) and charge (b) yields for $^{254}$Fm calculated with SkM$^*$ in the 2D (green solid lines) and 3D (blue dashed lines) collective spaces. Experimental yields  from~Refs.\,\cite{PhysRevC.8.1488,PhysRevC.16.1483} are shown as black and red circles respectively.}
    \label{fig:254-skms-yields}
\end{figure}

\begin{figure}
    \includegraphics[width=\linewidth]{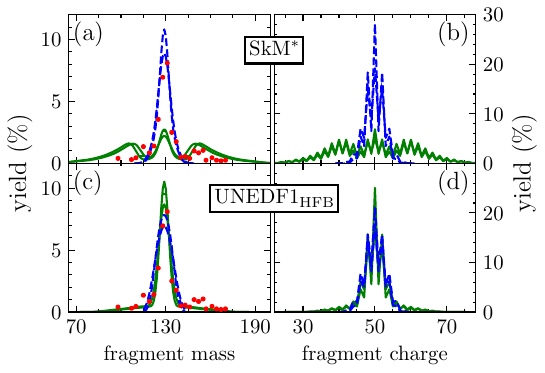}
    \caption{The fragment mass (left) and charge (right) yields for $^{258}$Fm using SkM$^{*}$ (top) and UNEDF1$_{\textrm{HFB}}$ (bottom) calculated in the 2D (green solid lines) and 3D (blue dashed lines) collective spaces. Experimental data from~\cite{PhysRevC.21.972} is shown as red filled circles.}
    \label{fig:3D-Fm258-yields}
\end{figure}

\begin{figure*}[htb]
    \includegraphics[width=1.0\linewidth]{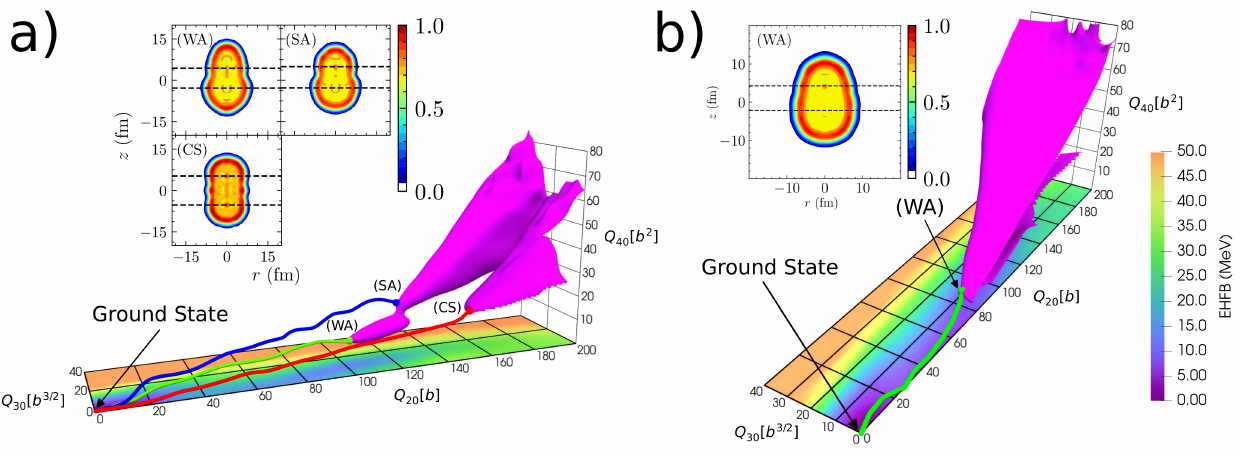} 
    \caption{Similar as in Fig.~\ref{fig:3d-Fm-surface} but for $^{306}$122. The 2D PESs are at constant $Q_{40} = 0$.}
    \label{fig:Z122-surface}
\end{figure*}

We first consider $^{254}$Fm as part of the benchmark chain. The identified modes and their relative probabilities for SkM$^*$ are consistent with those of~\cite{Staszczak2009}. Figure~\ref{fig:254-skms-yields} shows the mass and charge distributions for SkM$^{*}$ computed in both the 2D and 3D collective spaces. The yields are consistent, with minor changes in the yield peaks primarily due to changes in the PES  (and hence the exit point). Results for UNEDF1$_\textrm{HFB}$ are similar. So, fragment yields for individual fission modes are expected to be largely consistent between 2D and 3D collective spaces.

Next, we consider $^{258}$Fm. The outer turning surface, LAPs, and localizations at $\mathcal{C}$ are shown in Fig.~\ref{fig:3d-Fm-surface}(a)-(b) for SkM$^*$ and UNEDF1$_\textrm{HFB}$, respectively. The part of the path from g.s. to FI is not shown for clarity. Two symmetric modes are identified: the compact (CS) and elongated (ES) modes. $\mathcal{C}_\textrm{ES}$ has larger $Q_{40}$ than $\mathcal{C}_\textrm{CS}$, and hence the neck region for $\mathcal{C}_\textrm{ES}$ is thicker. We also found two asymmetric fission pathways: the weak (WA) and strong (SA) modes. $\mathcal{C}_\textrm{WA}$ has a small ($Q_{30}<10$~b$^{3/2}$) octupole moment. The ES mode is present due to the $Q_{40}$ constraint; the other modes are consistent with the discussion of Sec.~\ref{subsec:2d-Fm}. The modes are also largely consistent with the pathways identified in Refs.~\cite{Staszczak2009,Staszczak2011}.

\begin{table}[htb]
    \caption{Relative probabilities for the modes identified in the 3D collective space.}
    \begin{ruledtabular}
    \begin{tabular}{c|cc|cc}
          Mode & \multicolumn{2}{c|}{$^{258}$Fm} & \multicolumn{2}{c}{$^{306}$122}  \\
                   & SkM$^*$ & UNEDF1$_\textrm{HFB}$ & SkM$^*$ & UNEDF1$_\textrm{HFB}$\\
         \hline
         CS & $0.79$ & $0.47$ & $\sim$0 & $-$ \\
         ES & $\sim$0 & $\sim$0 & $-$ & $-$ \\
         WA & $0.21$ & $0.53$ & $1.0$ & $1.0$\\
         SA & $\sim$0 & $-$ & $\sim$0 & $-$\\
    \end{tabular}
    \end{ruledtabular}
    \label{table:3d-rel_action}
\end{table}

Table~\ref{table:3d-rel_action} lists the relative probability of each mode. The symmetric contribution is now larger (smaller) than in 2D for SkM$^*$ (UNEDF1$_\textrm{HFB}$). Figure~\ref{fig:3D-Fm258-yields} shows the corresponding fragment yields compared to the yields calculated in the 2D collective space. For UNEDF1$_\textrm{HFB}$, $\mathcal{C}_\textrm{WA}$ is found somewhat closer to the $Q_{30}=0$ axis than in 2D, resulting in a less asymmetric distribution. This worsens the agreement with the tails of the experimental yields. For SkM$^*$, agreement with experiment is improved due to the increase in the relative contribution of the symmetric mode. This highlights the importance of adding additional collective coordinates for mode identification.

\begin{figure}[htb]
    \includegraphics[width=\linewidth]{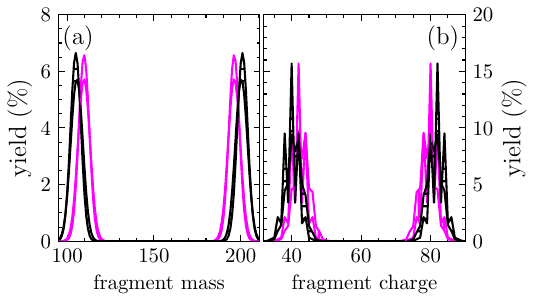}
    \caption{The fragment mass (a) and charge (b) yields for $^{306}$122 calculated with UNEDF1$_\textrm{HFB}$ (black solid lines) and SkM$^*$ (magenta dashed lines).}
    \label{fig:z-122-yields}
\end{figure}

Finally, in Fig.~\ref{fig:Z122-surface} we consider $^{306}$122, a superheavy nucleus that may exhibit multimodal fission~\cite{Staszczak2013,BARAN2015,sowmya2023competition,Mumpower2020,Kostryukov2021,Lemaitre2021,ishizuka2023nuclear}. A pronounced asymmetric valley forms beyond the outer turning surface leading to the cluster decay channel, similar to $^{294}$Og~\cite{Matheson2019}. The relative probabilities for both EDFs are listed in Table~\ref{table:3d-rel_action}, and the SkM$^*$ results are consistent with~\cite{Staszczak2013,BARAN2015}. The yields, shown in Fig.~\ref{fig:z-122-yields}, are in qualitative agreement between the EDFs. As with $^{294}$Og, the heavy fragment is close to the doubly magic nucleus $^{208}$Pb. The yields disagree with the symmetric yields calculated in the BSM model~\cite{Mumpower2020}.

\section{Conclusions}\label{sec:conclusions}
In this study, we have explored the competition between SF modes for $^{254,256,258,260}$Fm and several superheavy nuclei within the axial quadrupole-octupole and quadrupole-octupole-hexadecapole collective spaces. We used the nudged elastic band method~\cite{OurNEBPaper} to find SF modes and their relative probabilities, and computed fragment yields using the hybrid approach from Refs.~\cite{Sadhukhan2020,Sadhukhan2022}.

In general, identified fission modes are consistently predicted by the EDFs used. However, the relative probabilities of the modes were found to be fairly sensitive to the choice of EDF, leading to different overall fragment yields. In the 2D collective space, UNEDF1$_\textrm{HFB}$ was found to agree best with experiment. The differences in the fragment yields highlights the utility of multimodal fission when differentiating between models.

For $^{258}$Fm, an elongated symmetric fission mode has been identified in the 3D collective space, and the yields were found to be more symmetric than in the 2D space. Introducing the hexadecapole moment brought theoretical results closer to  experiment for SkM$^*$. The nucleus $^{306}122$ is predicted to decay via cluster emission, similar to $^{294}$Og~\cite{Matheson2019} and $^{284}$Cn shown in Fig.~\ref{fig:284Cn-yields}. Identification of a cluster decay of superheavy nuclei remains an important experimental challenge.

Coexistence of fission pathways is expected for many {\rpa} nuclei, and has a strong impact on the fragment yields. We demonstrated in this work that the nudged elastic band method~\cite{OurNEBPaper} is capable of efficiently identifying such competing pathways, enabling future studies of multimodal fission in several collective coordinates.

Furthermore, the above highlights the importance of including more collective coordinates when studying SF. The triaxial quadrupole moment, for instance, is important when describing the collective motion from the ground state to the fission isomer~\cite{Sheikh2009,Staszczak2011,Sadhukhan2013,Sadhukhan2014}. One can also consider a particle number dispersion term, which controls dynamic pairing fluctuations~\cite{Vaquero2011,Vaquero2013,Sadhukhan2014,Giuliani2014}. However, increasing the dimension of the collective space beyond the presented 3D calculations is computationally expensive. In this regard, DFT emulators will help by reducing  computational costs~\cite{Lasseri2020,Bonilla2022,duguet2023,Lay2023}.

One interesting alternative to the presented description of SF is dynamic reaction processes such as quasifission, which exhibits similar fragment production characteristics as fission and has been identified as a promising experimental avenue for studying the fission properties of hard-to-produce nuclei~\cite{godbey2019,simenel2021,McGlynn2023}. Time-dependent reaction methods do not constrain the system to a set of collective coordinates, allowing the most probable modes to be determined efficiently~\cite{godbey2020}. This efficiency enables systematic calculations across the nuclear chart, in addition to uncertainty quantification when describing real-time large amplitude collective motion beyond existing fusion studies~\cite{godbey2022}. Regardless of the method of exploration, comprehensive studies of multimodal fission are needed across the chart of nuclides. These systematic studies of fragment properties for spontaneous, neutron-induced, and compound-nucleus fission all contain rich information that will inform {\rpa} nucleosynthesis and provide key insights about the underlying structure and dynamics of exotic nuclei.

\PRLsep

\begin{acknowledgments} 
 This work was supported by the U.S. Department of Energy under Award Numbers DOE-DE-NA0004074 (NNSA, the Stewardship Science Academic Alliances program), DE-SC0013365 (Office of Science), and DE-SC0023175 (Office of Science, NUCLEI SciDAC-5 collaboration); by the National Science Foundation CSSI program under award number 2004601 (BAND collaboration); and by the Spanish Agencia Estatal de Investigaci{\'o}n (AEI) of the Ministry of Science and Innovation (MCIN) under grant agreements No.~PID2021-127890NB-I00 and No.~RYC2021-031880-I funded by MCIN/AEI/10.13039/501100011033 and the "European Union NextGenerationEU/PRTR".
\end{acknowledgments}

D.L. and E.F. contributed equally to this work.

\end{document}